\documentclass[aps,prd,twocolumn,showpacs,nofootinbib]{revtex4-1}

\usepackage{amsfonts,color,graphicx,epsfig,amsmath,multirow,slashed}
\usepackage[colorlinks=true,linkcolor=blue,citecolor=blue, urlcolor=blue]{hyperref}

\begin{document}

\title{Exclusive production of $J/\psi+\eta_c$ at the $B$ factories Belle and Babar using the principle of maximum conformality}

\author{Zhan Sun$^1$}
\email{zhansun@cqu.edu.cn}
\author{Xing-Gang Wu$^{2}$}
\email{wuxg@cqu.edu.cn}
\author{Yang Ma$^{3}$}
\email{mayangluon@pitt.edu}
\author{Stanley J. Brodsky$^{4}$}
\email{sjbth@slac.stanford.edu}

\affiliation{
\footnotesize
$^{1}$ Department of Physics, Guizhou Minzu University, Guiyang 550025, P. R. China. \\
$^{2}$ Department of Physics, Chongqing University, Chongqing 401331, P.R. China. \\
$^{3}$ PITT-PACC, Department of Physics and Astronomy, University of Pittsburgh, PA 15260, USA. \\
$^{4}$ SLAC National Accelerator Laboratory, Stanford University, Stanford, California 94039, USA. }

\date{\today}

\begin{abstract}

We predict the rate for exclusive double-charmonium production in electron-positron annihilation $e^+ e^- \to J/\psi+\eta_c$ using pQCD and the NRQCD framework for hard, heavy-quarkonium exclusive processes. The cross sections measured at the $B$-factories Belle and Babar at $\sqrt{s}=10.6$ GeV disagree with the pQCD leading-order predictions by an order of magnitude. The predictions at next-to-leading order are, however, very sensitive to the choice of the renormalization scale, resulting in an apparent discrepancy between the theoretical prediction and the data. We show that this discrepancy can in fact be eliminated by applying the Principle of Maximum Conformality (PMC) to set the renormalization scale. By carefully applying the PMC to different topologies of the annihilation process, one achieves precise pQCD predictions, together with improved perturbative convergence. We also observe that the single-photon-fragmentation QED correction is important, an effect which increases the total cross-section by about $10\%$. The scale-fixed, scheme-independent cross-section predicted by the PMC is $\sigma_{\rm tot}|_{\rm PMC}=20.35 ^{+3.5}_{-3.8}$ fb, where the uncertainties come from the squared average of the errors due to the value of the charm mass and the uncertainty from the quarkonium wavefunctions at the origin. We find that the typical momentum flow of the process is $2.30$ GeV, which explains the guessed choice of $2-3$ GeV using conventional scale-setting. The scale-fixed $e^+ e^- \to J/\psi+\eta_c$ cross-section predicted by the PMC shows agreement with the Belle and Babar measurements.


\end{abstract}

\maketitle

In the year 2002, the Belle Collaboration released the measurements on the total cross-section of the exclusive production of $J/\psi+\eta_c$ via the $e^{+}e^{-}$-annihilation at the center-of-mass collision energy $\sqrt{s}=10.58$ GeV~\cite{Abe:2002rb}. The result was $\sigma[e^+e^-\to J/\psi+\eta_c]\times{\mathcal B}_{\ge 4}=33_{-6}^{+7}\pm 9$ fb with ${\mathcal B}_{\ge 4}$ denoting the branching ratio of $\eta_c$ into four or more charged tracks. This measurements were afterwards improved as $\sigma[e^+e^- \to J/\psi+\eta_c] \times {\mathcal B}_{>2} = 25.6\pm 2.8\pm3.4$ fb \cite{Abe:2004ww}. Later in the year 2005, the BaBar Collaboration independently measured the total cross-section and obtained $17.6\pm2.8^{+1.5}_{-2.1}$ fb \cite{Aubert:2005tj}. However, the leading-order (LO) calculation derived using the framework of the nonrelativistic QCD (NRQCD) factorization theory \cite{Bodwin:1994jh}, gives a prediction for the total cross-section $\sim 2.3-5.5$ fb \cite{Braaten:2002fi, Liu:2002wq, Hagiwara:2003cw, Liu:2004ga}, an order of magnitude smaller than the measured value.

A number of theoretical attempts have been suggested in order to explain this discrepancy, either by including the $\mathcal O(\alpha_s)$-, $\mathcal O(v^2)$- and $\mathcal O(\alpha_s v^2)$- corrections to the NRQCD prediction, by using the light-cone factorization approach, or by using QCD light-cone sum rules, cf. Refs.\cite{Ma:2004qf, Bondar:2004sv, Braguta:2005kr, Bodwin:2006dm, Sun:2009zk, Zhang:2005cha, Gong:2007db}. The QCD next-to-leading-order (NLO) calculation of the process $e^+e^-\to J/\psi+\eta_c$ \cite{Zhang:2005cha, Gong:2007db} has been regarded as a breakthrough, showing that the NLO-terms are large and positive. The Belle and Babar data could thus be explained by choosing the renormalization scale $\mu_{r}\sim 2-3$ GeV. However, the NLO prediction is highly sensitive to the choice of $\mu_{r}$; for example, as will be shown below, the total cross-section is decreased by about $50\%$ by varying $\mu_{r}$ from $3.0$ to $10.6$ GeV. Thus one cannot draw any definite conclusion on the validity of the theoretical approach.

A renormalization scale ambiguity apparently exists in any fixed-order pQCD prediction; it is usually regarded as an important systematic error for the pQCD predictions~\cite{Wu:2013ei, Wu:2014iba}. It is thus crucial to eliminate this scale uncertainty in order to achieve a definitive prediction. The {\it Principle of Maximum Conformality } (PMC)~\cite{Brodsky:2011ta, Mojaza:2012mf, Brodsky:2013vpa, Brodsky:2012rj} provides a rigorous scale-setting approach which is independent of the choice of the renormalization scheme, such as the modified-minimal-subtraction ($\overline{\mathrm{MS}}$) scheme~\cite{Bardeen:1978yd}. It provides the underlying reason for the Brodsky-Lepage-Mackenzie (BLM) scale-setting method~\cite{Brodsky:1982gc} and extends it to all orders in pQCD. In this paper we will show how one can apply the PMC formalism to the $e^+ e^- \to J/\psi+\eta_c$ cross section and to thus investigate whether the disagreement between theory and experiment can be eliminated.

\begin{widetext}
\begin{center}
\begin{figure}[htb]
\includegraphics[width=0.9\textwidth]{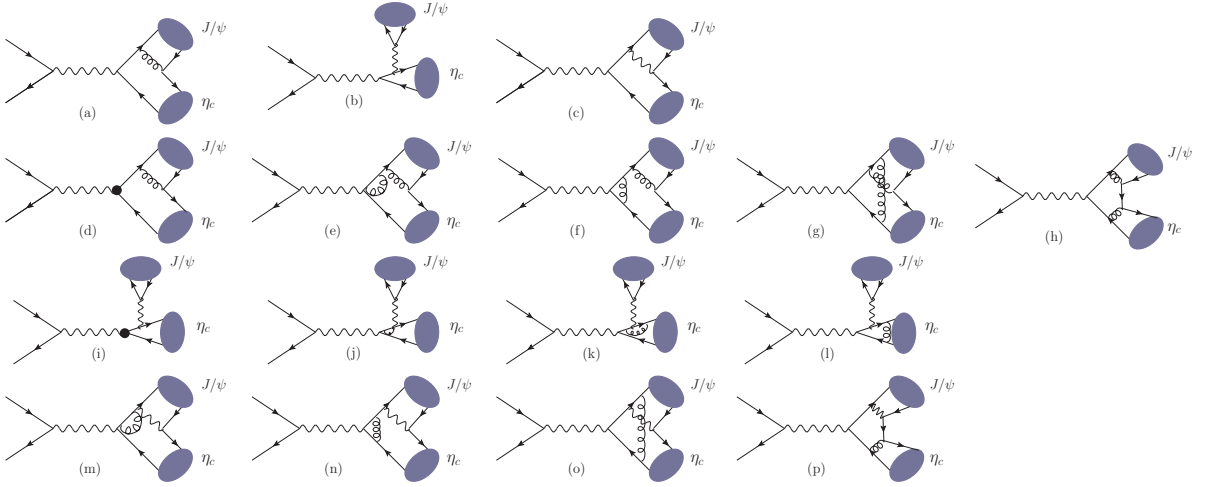}
\caption{Typical Feynman Diagrams for $e^+e^- \to J/\psi+\eta_c$ up to ${\cal O}(\alpha^3 \alpha_s^3)$-order level. Fig.(\ref{fig:Feyn}a) is the QCD LO diagram and Figs.(\ref{fig:Feyn}b-\ref{fig:Feyn}c) are the QED LO diagrams, in which Fig.(\ref{fig:Feyn}b) denotes the typical SPF (single-photon-fragmentation) diagram. Figs.(\ref{fig:Feyn}d-\ref{fig:Feyn}h) are NLO QCD corrections to Fig.(\ref{fig:Feyn}a) and Figs.(\ref{fig:Feyn}i-\ref{fig:Feyn}p) are NLO QCD corrections to Figs.(\ref{fig:Feyn}b-\ref{fig:Feyn}c).}
\label{fig:Feyn}
\end{figure}
\end{center}
\end{widetext}

The squared matrix element of $e^{+}e^{-} \to J/\psi+\eta_c$ can be schematically written as,
\begin{eqnarray}
&& |(\mathcal M_{\alpha\alpha_s}+\mathcal M_{\alpha\alpha_s^2})+(\mathcal M_{\alpha^2}+\mathcal M_{\alpha^2\alpha_s})|^2 \nonumber \\
 &=&|\mathcal M_{\alpha\alpha_s}|^2+2\textrm{Re}(\mathcal M_{\alpha\alpha_s}\mathcal M_{\alpha\alpha_s^2}^{*}) +2\textrm{Re}(\mathcal M_{\alpha\alpha_s}\mathcal M_{\alpha^2}^{*}) \nonumber \\
   &&+2\textrm{Re}(\mathcal M_{\alpha\alpha_s}\mathcal M_{\alpha^2\alpha_s}^{*}) +2\textrm{Re}(\mathcal M_{\alpha\alpha_s^2}\mathcal M_{\alpha^2}^{*})+\cdots
\end{eqnarray}
There is no contribution from real gluons and photons. There are in total 4 QCD diagrams (4 tree-level, 60 one-loop and 20 counter-terms) and 72 QED diagrams (6 tree-level, 42 one-loop and 24 counter-terms) for $e^+e^- \to J/\psi+\eta_c$. Typical Feynman diagrams are illustrated in Fig.(\ref{fig:Feyn}). We will not need to calculate the NLO QED correction to the diagrams such as Fig.(\ref{fig:Feyn}a) in order to calculate the amplitude $M_{\alpha^2\alpha_s}$, since these topologies are compensated by the initial-state radiation diagrams. Such contributions are irrelevant to the exclusive $e^{+}e^{-}$ annihilation processes.

Fig.(\ref{fig:Feyn}a), with three other permutation diagrams, form a gauge-invariant subset and contribute to the LO squared matrix element $|M_{\alpha\alpha_s}|^2$. Fig.(\ref{fig:Feyn}b) together with another diagram formed by attaching the second (virtual) photon to the anti-charm quark line are single-photon-fragmentation (SPF) diagrams which also form a gauge-invariant subset and contribute to the cross-term $2\textrm{Re}(\mathcal M_{\alpha\alpha_s}\mathcal M_{\alpha^2}^{*})$. Their contributions suffer from the usual $e_c^2\alpha/\alpha_s$-suppression but have a large kinematic enhancement, leading to a sizable $\sim 20\%$ contribution to the LO QCD cross-section~\cite{Braaten:2002fi, Liu:2004ga}. Fig.(\ref{fig:Feyn}c) with three permutation diagrams also suffer from the $e_c^2\alpha/\alpha_s$-suppression, giving contributions of order of $1\%$ of the LO QCD cross-section.

In the following NLO calculation, we will also calculate the NLO-contributions of significant QED diagrams which so far have not been considered in the literature. Figs.(\ref{fig:Feyn}d-\ref{fig:Feyn}h) are typical NLO contributions to the LO QCD diagrams which contribute to the cross-term $2\textrm{Re}(\mathcal M_{\alpha\alpha_s}\mathcal M_{\alpha\alpha_s^2}^{*})$ at the $\alpha^2\alpha_s^3$-order level. Figs.(\ref{fig:Feyn}i-\ref{fig:Feyn}p) are typical NLO contributions to the QED diagrams, which contribute to the cross-term $2\textrm{Re}(\mathcal M_{\alpha\alpha_s}\mathcal M_{\alpha^2\alpha_s}^{*})$ and $2\textrm{Re}(\mathcal M_{\alpha\alpha_s^2}\mathcal M_{\alpha^2}^{*})$ at the $\alpha^3\alpha_s^2$-order level. In doing the numerical calculation, we observe that that the contributions of the SPF diagrams are dominant over that of the remaining QED diagrams.

We divide the differential cross-section into the following four parts:
\begin{equation}
d\sigma = d\sigma_{\alpha^2}^{(0)}+d\sigma_{\alpha^2}^{(1)}+d\sigma_{\alpha^3}^{(0)}+d\sigma_{\alpha^3}^{(1)}
\end{equation}
with
\begin{eqnarray}
d\sigma_{\alpha^2}^{(0)} &\propto& |\mathcal M_{\alpha\alpha_s}|^2, \\
d\sigma_{\alpha^2}^{(1)} &\propto& 2\textrm{Re}(\mathcal M_{\alpha\alpha_s}\mathcal M_{\alpha\alpha_s^2}^{*}), \\
d\sigma_{\alpha^3}^{(0)} &\propto& 2\textrm{Re}(\mathcal M_{\alpha\alpha_s}\mathcal M_{\alpha^2}^{*}),  \\
d\sigma_{\alpha^3}^{(1)} &\propto& 2\textrm{Re}(\mathcal M_{\alpha\alpha_s}\mathcal M_{\alpha^2\alpha_s}^{*})+2\textrm{Re}(\mathcal M_{\alpha^2}\mathcal M_{\alpha\alpha_s^2}^{*}).
\end{eqnarray}
The first two terms $d\sigma_{\alpha^2}^{(0,1)}$ and the second two terms $d\sigma_{\alpha^3}^{(0,1)}$ are the usual QCD contributions up to NLO level and the new QED contributions up to NLO level, respectively.

In order to eliminate the ultraviolet (UV) and infrared (IR) divergences, we will adopt the usual dimensional renormalization procedure with $D=4-2\epsilon$. The on-mass-shell (OS) scheme is employed to set the renormalization constants of the charm-quark mass $Z_m$ and the filed $Z_2$, and the $\overline{\rm MS}$-scheme for the QCD gauge coupling $Z_g$ and the gluon field $Z_3$~\cite{Zhang:2005cha}\footnote{For the QCD correction to the QED LO diagrams, only $Z_{m}$ and $Z_2$ are involved.},
\begin{eqnarray}
\delta Z_{m}^{\rm OS}&=& -3 C_{F} \frac{\alpha_s N_{\epsilon}}{4\pi}\left[\frac{1}{\epsilon_{\textrm{UV}}}-\gamma_{E}+\textrm{ln}\frac{4 \pi \mu_r^2}{m_c^2}+\frac{4}{3}+\mathcal O(\epsilon)\right], \nonumber \\
\delta Z_{2}^{\rm OS}&=& - C_{F} \frac{\alpha_s N_{\epsilon}}{4\pi}\left[\frac{1}{\epsilon_{\textrm{UV}}}+\frac{2}{\epsilon_{\textrm{IR}}}-3 \gamma_{E}+3 \textrm{ln}\frac{4 \pi \mu_r^2}{m_c^2} \right. \nonumber\\
&& \left.+4+\mathcal O(\epsilon)\right], \nonumber \\
\delta Z_{3}^{\overline{\rm MS}}&=& \frac{\alpha_s N_{\epsilon}}{4\pi}(\beta_{0}-2 C_{A})\left[\frac{1}{\epsilon_{\textrm{UV}}}-\gamma_{E}+\textrm{ln}(4\pi)+\mathcal O(\epsilon)\right], \nonumber \\
\delta Z_{g}^{\overline{\rm MS}}&=& -\frac{\beta_{0}}{2}\frac{\alpha_s N_{\epsilon}}{4\pi}\left[\frac{1} {\epsilon_{\textrm{UV}}}-\gamma_{E}+\textrm{ln}(4\pi)+\mathcal O(\epsilon)\right],
\end{eqnarray}
where $\gamma_E$ is the Euler's constant, $\beta_{0}=\frac{11}{3}C_A-\frac{4}{3}T_Fn_f$ is the one-loop coefficient of the $\beta$-function and $n_f$ is the active quark flavor numbers, $N_{\epsilon}= \Gamma[1-\epsilon] /({4\pi\mu_r^2}/{(4m_c^2)})^{\epsilon}$. In ${\rm SU}(3)_c$, the color factors are given by $T_F=\frac{1}{2}$, $C_F=\frac{4}{3}$ and $C_A=3$. It is noted that the renormalization scale $\mu_r$ occurs via a unique form of $\beta_0 \textrm{ln}({m_c^2}/{\mu_r^2})$ in the present NLO pQCD series, and one can unambiguously apply the PMC to set the optimal renormalization scale. Although we will utilize dimensional regularization, the final PMC prediction is independent of the choice of the renormalization scheme.

We adopt the package \textbf{Malt@FDC}~\cite{Wang:2004du, Feng:2017bdu, Sun:2017nly, Sun:2017wxk} to do the NLO calculation. As a cross-check of our calculation, we have verified that by taking the same input parameters, we obtain the same NLO predictions for $\sigma^{(0)}_{\alpha^2}$ and $\sigma^{(1)}_{\alpha^2}$ as those of Refs.\cite{Zhang:2005cha, Gong:2007db}.

In order to do the numerical calculation, we will assume $m_c=1.5$ GeV, $M_{J/\psi}=M_{\eta_c}=2m_c$, and $\alpha=1/137$. The $e^+e^-$-collision energy is assumed to be $\sqrt{s}=10.6$ GeV. The two-loop pQCD prediction for the $\alpha_s$ running coupling is assumed.
\begin{eqnarray}
\alpha_s(\mu_{r})=\frac{4\pi}{\beta_0 L}-\frac{4\pi\beta_1 \ln L} {\beta_0^3 L^2}, \label{runningcouple1}
\end{eqnarray}
where $L=\ln(\mu^2_{r}/\Lambda^2_{\rm QCD})$ with $\Lambda^{(4)}_{\overline{\rm MS}}=0.332$ GeV which is fixed by requiring $\alpha_s(M_Z)=0.118$. The $J/\psi$ and $\eta_c$ wavefunctions at the origin are taken as $|R_s^{\eta_c}(0)|^2 = |R_s^{J/\psi}(0)|^2=0.978~\textrm{GeV}^3$~\cite{Zhang:2005cha}.

\begin{table}[htb]
\centering
\caption{Total cross-section for $e^{+}e^{-} \to J/\psi+\eta_c$ under conventional scale-setting (in unit: fb). Three typical renormalization scale $\mu_r$ are adopted. $m_c$=1.5 GeV. }
\label{table:cross section1}
\begin{tabular}{ccccccccccc}
\hline
 & $\sigma_{\alpha^2}^{(0)}$ & $\sigma_{\alpha^2}^{(1)}$ & $\sigma_{\alpha^3}^{(0)}$ & $\sigma_{\alpha^3}^{(1)}$ & $\sigma_{\rm tot}$ & $K_{\alpha^2}$ & $K_{\alpha^3}$ & $r_{\alpha^2}$ & $r_{\alpha^3}$ \\ \hline
 $\mu_r=3$ GeV & $7.83$ & $7.58$ & $1.53$ & $0.19$ & $17.13$ & $1.97$ & $1.12$ & $90\%$ & $10\%$ \\
 $\mu_r=6$ GeV & $4.92$ & $5.63$ & $1.22$ & $0.34$ & $12.11$ & $2.14$ & $1.28$ & $87\%$ & $13\%$ \\
 $\mu_r=10$ GeV & $3.80$ & $4.73$ & $1.07$ & $0.40$ & $10.00$ & $2.25$ & $1.37$ & $85\%$ & $15\%$ \\ \hline
\end{tabular}
\end{table}

\begin{figure}[htb]
\includegraphics[width=0.45\textwidth]{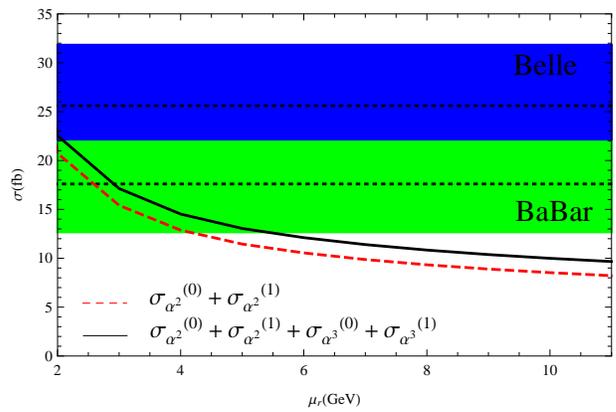}
\caption{The renormalization scale ($\mu_r$) dependence of the total cross-section using the conventional scale-setting. The dashed line denotes the pure QCD contributions up to NLO level and the solid line is the total cross-section including the QED contributions up to NLO level. The Belle~\cite{Abe:2004ww} and Babar~\cite{Aubert:2005tj} measurements are presented as a comparison, which are shown by the shaded bands. }
\label{fig:conv}
\end{figure}

The total cross-section predicted using conventional scale setting with three typical choices of the renormalization scale are shown in Table \ref{table:cross section1}. Here $\sigma_{\rm tot}=\sigma_{\alpha^2}^{(0)}+\sigma_{\alpha^2}^{(1)}+ \sigma_{\alpha^3}^{(0)}+\sigma_{\alpha^3}^{(1)}$ and the ratios $K_{\alpha^2}$, $K_{\alpha^3}$, $r_{\alpha^2}$, $r_{\alpha^3}$ are defined as ${(\sigma_{\alpha^2}^{(0)}+\sigma_{\alpha^2}^{(1)})}/{\sigma_{\alpha^2}^{(0)}}$, ${(\sigma_{\alpha^3}^{(0)} +\sigma_{\alpha^3}^{(1)})}/{\sigma_{\alpha^3}^{(0)}}$, ${(\sigma_{\alpha^2}^{(0)}+\sigma_{\alpha^2}^{(1)})}/{\sigma_{\rm tot}}$, ${(\sigma_{\alpha^3}^{(0)} +\sigma_{\alpha^3}^{(1)})}/{\sigma_{\rm tot}}$, respectively.
We note that the QED (especially the SPF) diagrams up to NLO level can enhance the pure QCD results by about $10\%$-$15\%$, thus their contributions should be taken into consideration.

The total cross-section exhibits a strong dependence on the renormalization scale $\mu_r$; it decreases from $17.13$ to $10$ fb by varying $\mu_r$ from $3$ to $10$ GeV. This strong scale dependence can also be seen in Fig.\ref{fig:conv}. By taking a smaller scale $\sim 2-3$ GeV, one obtains a larger cross-section in agreement with the Belle and Babar measurements. However if one can only guess the scale, the predictive power of NRQCD theory is lost, and one cannot draw any definite conclusions on the validity of the pQCD prediction. It is thus crucial to eliminate the renormalization scale dependence.

The PMC provides a systematic way to eliminate the renormalization scale uncertainty for the pQCD predictions. Fig.\ref{fig:Feyn} shows there are two types of Feynman diagrams, i.e. the ones with or without the QED diagrams, indicating their typical momentum flows may be different. To apply the PMC, we divide the total cross-section into two parts, $\sigma_{\alpha^2}=\sigma_{\alpha^2}^{(0)}+\sigma_{\alpha^2}^{(1)}$ and $\sigma_{\alpha^3}=\sigma_{\alpha^3}^{(0)} +\sigma_{\alpha^3}^{(1)}$, which can be schematically rewritten as
\begin{eqnarray}
\sigma_{\alpha^2}&=&A_1 \alpha^2_{s,\overline{\rm MS}}(\mu_r)\left[1+\frac{\alpha_{s,\overline{\rm MS}}(\mu_r)}{\pi}(B_1 n_f + C_1) \right] \nonumber \\
&=& A_1 \alpha^2_{s,\overline{\rm MS}}(Q_1)\left[1+ D^*_{1}\frac{\alpha_{s,\overline{\rm MS}}(Q_1)}{\pi} \right] \label{eqalp2}
\end{eqnarray}
and
\begin{eqnarray}
\sigma_{\alpha^3}&=&A_2 \alpha_{s,\overline{\rm MS}}(\mu_r)\left[1+\frac{\alpha_{s,\overline{\rm MS}}(\mu_r)}{\pi}(B_2 n_f + C_2) \right] \nonumber \\
&=& A_2 \alpha_{s,\overline{\rm MS}}(Q_2)\left[1+ D^*_{2}\frac{\alpha_{s,\overline{\rm MS}}(Q_2)}{\pi} \right]. \label{eqalp3}
\end{eqnarray}
The second equations in Eqs.(\ref{eqalp2}, \ref{eqalp3}) are obtained by using the standard PMC procedures, and the PMC scales
\begin{eqnarray}
Q_1 &=& \frac{1}{2}\mu_r e^{3B_1}, \;\; Q_2 = \mu_r e^{3B_2},
\end{eqnarray}
and the conformal coefficients
\begin{eqnarray}
D^{*}_1 &=& \frac{33}{2} B_1 + C_1, \;\; D^{*}_2 = \frac{33}{2} B_2 + C_2.
\end{eqnarray}
Numerical values for the coefficients $A_i$, $B_i$ and $C_i$ can be obtained by using the package \textbf{Malt@FDC}.

It is interesting to find that $Q_1=Q_2\equiv2.30$ GeV for any choice of initial renormalization $\mu_r$ \footnote{A strict demonstration of such scale-independence for the leading-order PMC scale has been given in Ref.\cite{Brodsky:2013vpa}. }, i.e. the PMC scales for the two different topologies are exactly the same. This can be explained by the fact that the $\beta_0$-terms are the same for both topologies, and thus the running behavior of the coupling constant determined by RGE are identical. The PMC scales $Q_{1,2}$ are independent of the choice of the initial choice of renormalization scale $\mu_r$; thus the conventional renormalization scale ambiguities are completely removed. The values of the PMC scales also show that the usually guessed value of 2-3 GeV, is in fact, correct.

\begin{table}[htb]
\centering
\caption{Total cross-section for $e^{+}e^{-} \to J/\psi+\eta_c$ using PMC scale-setting (in unit: fb). The initial renormalization scale $\mu_r\in[3,10]$ GeV. $m_c$=1.5 GeV. }
\label{table:cross section2}
\begin{tabular}{ccccccccccc}
\hline
 & $\sigma_{\alpha^2}^{(0)}$ & $\sigma_{\alpha^2}^{(1)}$ & $\sigma_{\alpha^3}^{(0)}$ & $\sigma_{\alpha^3}^{(1)}$ & $\sigma_{\rm tot}$ & $K_{\alpha^2}$ & $K_{\alpha^3}$ & $r_{\alpha^2}$ & $r_{\alpha^3}$ \\
 \hline
 $\mu_r\in[3,10]$GeV & $9.86$ & $8.71$ & $1.72$ & $0.06$ & $20.35$ & $1.88$ & $1.03$ & $91\%$ & $9\%$ \\
\hline
\end{tabular}
\end{table}

The prediction for the total cross-sections after applying the PMC, taking the initial renormalization scale in the range $\mu_r \in[3,10]$ GeV, are shown in Table \ref{table:cross section2}. Table \ref{table:cross section2} shows that the renormalization scale uncertainty is eliminated by applying the PMC. Moreover, one observes that the PMC scales are independent of the choice of the initial scale $\mu_r$. The PMC prediction is also scheme independent, which can be confirmed by the commensurate scale relations among different observables~\cite{Brodsky:1994eh}. A demonstration of the renormalization scheme independence for the PMC prediction has been given in Ref.\cite{Wu:2018cmb}. The convergence of the cross-section $\sigma_{\alpha^3}$ has been greatly improved, leading to a much smaller $K_{\alpha^3}(=1.03)$ factor. In contrast with other PMC predictions, the convergence of the cross-section $\sigma_{\alpha^2}$ ($K_{\alpha^2}=1.88$) remains an issue because of the large conformal coefficient $D^*_1$.

\begin{figure}[htb]
\includegraphics[width=0.45\textwidth]{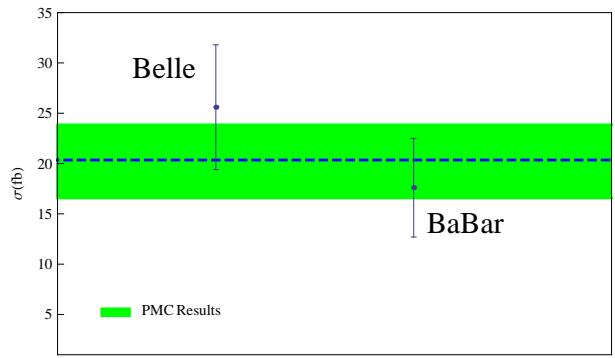}
\caption{\label{fig:pmc}
Our final PMC prediction for the total cross-section of $e^{+}e^{-} \to J/\psi+\eta_c$. The shaded band denotes the squared average errors of the uncertainties from $m_c=1.5 \pm 0.1$ GeV and the wavefunctions at the zero $|R_s^{\eta_c}(0)|^2 = |R_s^{J/\psi}(0)|^2=0.978\pm 0.04~\textrm{GeV}^3$, whose central value is for $m_c=1.5$ GeV and $|R_s^{\eta_c}(0)|^2 = |R_s^{J/\psi}(0)|^2=0.978~\textrm{GeV}^3$. The Belle~\cite{Abe:2004ww} and Babar~\cite{Aubert:2005tj} measurements are presented as a comparison. }
\end{figure}

Our final prediction after applying the PMC, is presented in Fig.\ref{fig:pmc}. The error band is obtained by taking $m_c=1.5\pm0.1$ GeV and $|R_s^{\eta_c}(0)|^2 = |R_s^{J/\psi}(0)|^2=0.978\pm 0.04~\textrm{GeV}^3$, where the uncertainties due to the wavefunctions at the origin reflect the constraint $\Gamma[J/\psi \to e^{+}e^{-}]=5.4 \pm 0.15 \pm 0.07$ GeV~\cite{Eidelman:2004wy}. One finds that the PMC provides a precise scale-fixed NRQCD prediction, in agreement with the Belle~\cite{Abe:2004ww} and Babar~\cite{Aubert:2005tj} measurement.

Within the framework of NRQCD, the color-octet components may also have sizable contributions. As an estimate, we calculate the color-octet channels, such as $e^{+}e^{-} \to {^{3}S_{1}^{[8]}+^{1}S_{0}^{[8]}} $, $e^{+}e^{-} \to {^{3}S_{1}^{[8]}+^{3}P_{J}^{[8]}}$, $e^{+}e^{-} \to {^{1}S_{0}^{[8]}+^{1}P_{1}^{[8]}}$ and $e^{+}e^{-} \to {^{1}P_{1}^{[8]}+^{3}P_{J}^{[8]}}$, by using the color-octet matrix elements derived under the heavy quark spin symmetry~\cite{Zhang:2014ybe, Sun:2015pia, Chao:2012iv, Bodwin:2014gia}. We find that the total color-octet contributions are negligibly small -- only about $0.1$ fb, for the present process.

In summary, by applying the PMC and taking contributions from the relevant QED (especially the SPF) diagrams into consideration, we have calculated the exclusive production channel $e^+e^- \to J/\psi+\eta_c$ up to NLO level. Our calculation shows that the SPF diagrams are important and increase the total cross-section by about $10\%$. The PMC result shows that the typical momentum flow of the process is $2.3$ GeV. After applying the PMC, the total cross-section increases to $\sigma_{\rm tot}|_{\rm PMC}=20.35 ^{+3.5}_{-3.8}$ fb, in which the uncertainty is the squared average of the errors from the charm quark mass and the wavefunctions at the origin. The PMC prediction agrees with the measurements of the Belle and Babar experiments within errors. This successful application of the PMC illustrates the importance of correct, rigorous, renormalization scale-setting; it also supports the applicability of NRQCD to hard exclusive processes involving heavy quarkonium.

\hspace{1cm}

\noindent{\bf Acknowledgments}:
We thank Hong-Fei Zhang, Sheng-Quan Wang and Hai-Bing Fu for helpful discussions. This work are supported in part by the Natural Science Foundation of China under the Grant No.11705034 and No.11625520, by the Department of Energy Contract No. DE-AC02-76SF00515, by the Project for Young Talents Growth of Guizhou Provincial Department of Education under Grant No.KY[2017]135 and the Key Project for Innovation Research Groups of Guizhou Provincial Department of Education under Grant No.KY[2016]028, and by the Fundamental Research Funds for the Central Universities under the Grant No.2018CDPTCG0001/3. SLAC-PUB-17300. PITT-PACC-1812. \\

\end{document}